\documentclass[rnote]{aa} 

\usepackage{graphicx}
\usepackage{txfonts, sidecap}
\usepackage{natbib,twoopt}
\usepackage[breaklinks=true]{hyperref} 
\bibpunct{(}{)}{;}{a}{}{,}             
\makeatletter
  \newcommandtwoopt{\citeads}[3][][]{\href{http://adsabs.harvard.edu/abs/#3}%
    {\def\hyper@linkstart##1##2{}%
     \let\hyper@linkend\@empty\citealp[#1][#2]{#3}}}
  \newcommandtwoopt{\citepads}[3][][]{\href{http://adsabs.harvard.edu/abs/#3}%
    {\def\hyper@linkstart##1##2{}%
     \let\hyper@linkend\@empty\citep[#1][#2]{#3}}}
  \newcommandtwoopt{\citetads}[3][][]{\href{http://adsabs.harvard.edu/abs/#3}%
    {\def\hyper@linkstart##1##2{}%
     \let\hyper@linkend\@empty\citet[#1][#2]{#3}}}
  \newcommandtwoopt{\citeyearads}[3][][]%
    {\href{http://adsabs.harvard.edu/abs/#3}
    {\def\hyper@linkstart##1##2{}%
     \let\hyper@linkend\@empty\citeyear[#1][#2]{#3}}}
\makeatother

\begin{document}

   \title{Disentangling the jet emission from protostellar systems}

   \subtitle{The ALMA view of VLA1623}

   \author{G. Santangelo
          \inst{1}
          \and
          N.~M. Murillo
          \inst{2}
          \and
          B. Nisini
          \inst{1}
          \and
          C. Codella
          \inst{3}
          \and
          S. Bruderer
          \inst{2}
          \and
          S.-P. Lai
          \inst{4,5}
          \and
          E.~F. van Dishoeck
          \inst{2,6}
          }

   \institute{Osservatorio Astronomico di Roma, via di Frascati 33, 
              00040 Monteporzio Catone, Italy\\
              \email{gina.santangelo@oa-roma.inaf.it}
         \and         
              Max Planck Institut f{\"u}r Extraterrestrische Physik (MPE), Giessenbachstr.1, 85748 Garching, Germany
         \and
              Osservatorio Astrofisico di Arcetri, Largo Enrico Fermi 5, I-50125 Florence, Italy
         \and
              Institute of Astronomy and Department of Physics, National Tsing Hua University, 101 Section 2 Kuang Fu Road, 30013 Hsinchu, Taiwan
         \and
              Academia Sinica Institute of Astronomy and Astrophysics, PO Box 23-141, 10617 Taipei, Taiwan         
         \and
              Leiden Observatory, Leiden University, P.O. Box 9513, 2300 RA Leiden, the Netherlands
             }

   \date{Received April 28, 2015; accepted June 16, 2015}

 
  \abstract
   {High-resolution studies of class 0 protostars represent the key to constraining protostar formation models. 
VLA16234$-$2417 represents the prototype of class 0 protostars, and it has been recently identified as a triple non-coeval system.
   }
   {We aim at deriving the physical properties of the jets in VLA16234$-$2417 
using tracers of shocked gas.}
   {ALMA Cycle 0 Early Science observations of CO(2$-$1) in the extended configuration are presented 
in comparison with previous SMA CO(3$-$2) and {\it Herschel}-PACS [O{\sc i}] 63~$\mu$m observations. 
Gas morphology and kinematics were analysed to constrain the physical structure and origin of the protostellar outflows.
   }
   {We reveal a collimated jet component associated with the [O{\sc i}] 63~$\mu$m emission 
at about 8$^{\prime\prime}$ ($\sim$960~AU) from source B. 
This newly detected jet component is inversely oriented with respect to the large-scale outflow driven by source A, and 
it is aligned with compact and fast jet emission very close to source B (about 0$.\!\!^{\prime\prime}$3)
rather than with the direction perpendicular to the A disk.
We also detect a cavity-like structure at low projected velocities, which surrounds the [O{\sc i}] 63~$\mu$m emission 
and is possibly associated with the outflow driven by source A.
Finally, no compact outflow emission is associated with source W.
   }
   {Our high-resolution ALMA observations seem to suggest there is a 
fast and collimated jet component associated with source B. This scenario would confirm that source B is younger than A, 
that it is in a very early stage of evolution, and that it drives a faster, more collimated, and more compact jet with respect to the large-scale slower 
outflow driven by A. 
However, a different scenario of a precessing jet driven by A cannot be firmly excluded from the present observations.
   }



   \maketitle
%

\section{Introduction}

With lifetimes less than 10$^5$ yr, class 0 objects represent 
the earliest phase of star formation, when most of their mass is still in the form 
of dense envelopes \citep[e.g.][]{andre2000,evans2009}.
Because they have not undergone significant evolution, 
they are likely to retain information on the physical and chemical initial conditions 
and on the physics of the collapse phase.
High-resolution studies of class 0 protostars thus represent the key 
to constraining protostar formation models.

Located in $\rho$ Ophiuchus \citep[$d$$\sim$120 pc,][]{loinard2008}, 
VLA16234$-$2417 (hereafter VLA1623) is considered to be the prototype of class 0 protostars \citep{andre1990,andre1993}.
It was originally identified as a strong continuum source at the mm wavelengths 
associated with a large-scale bipolar outflow traced by CO and H$_2$ emissions \citep{andre1990,dent1995,yu1997}
nearly on the plane of the sky \citep[$i$$=$75$^{\circ}$,][]{davis1999}.
High-resolution observations by \citet{bontemps1997} revealed multiple continuum sources, which 
were interpreted as a radio jet driving the large-scale outflow 
rather than a multiple protostellar system \citep[see also][]{maury2010,maury2012}.
However, VLA1623's multiplicity \citep{looney2000,dent1995,yu1997} 
was recently confirmed by \citet{murillo2013a} and \citet{chen2013} through SMA observations.
In particular, \citet{murillo2013a} show that VLA1623 represents a triple non-coeval system composed of
VLA1623A, VLA1623B, and VLA1623W (hereafter A, B, and W), with each source driving its own outflow. 
Source A was identified as a deeply embedded class 0 source with no emission at wavelengths shorter than 24~$\mu$m. 
It shows a rotationally supported disk in C$^{18}$O and C$^{17}$O, later confirmed through ALMA 
observations by \citet{murillo2013b}, 
and drives the large-scale bipolar outflow. It has been suggested that
source B is a cold and compact source in a very early stage of star formation 
between the starless core and class 0 stages with a separation of 1$.\!\!^{\prime\prime}$1 ($\sim$132~AU) from A 
\citep[see also][]{looney2000,maury2012}. \citet{murillo2013a} suggest that source B drives a pole-on outflow.
 Finally,  it has been argued that source W is an $L$$=$1.4 $L_{\odot}$ 
class I source at the projected distance of 10$^{\prime\prime}$ from A.

   \begin{figure*}
   \centering
   \includegraphics[width=0.98\textwidth]{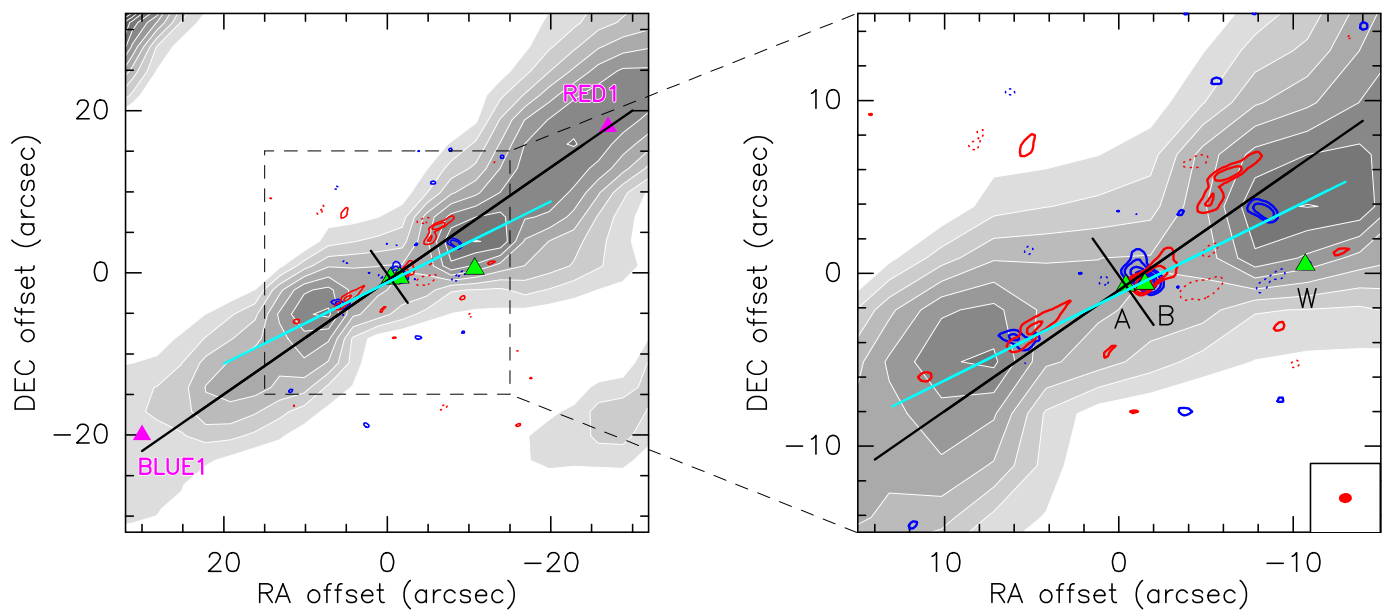}
   \caption{ALMA CO(2$-$1) emission integrated over the blue- ($-$8, $+$1 km s$^{-1}$) and red-shifted ($+$5, $+$18 km s$^{-1}$) 
velocity ranges are compared with the PACS [O{\sc i}] emission line at 63~$\mu$m by \citet{nisini2015}.
The ALMA synthesised beam of the CO map is shown in the right panel.
Green triangles indicate the positions of the three sources, as reported by \citet{murillo2013b}; 
magenta triangles label the closest CO knots identified by \citet{andre1990}. 
Black lines indicate the directions of the VLA1623A disk \citep[PA=35$^{\circ}$,][]{murillo2013a,murillo2013b} 
and the relative outflow (PA=125$^{\circ}$). The latter is consistent with the large-scale CO outflow direction 
and with the secondary [O{\sc i}] 63~$\mu$m peaks, whereas the closest [O{\sc i}] 63~$\mu$m emission peaks are misaligned.
The cyan line indicates the direction of the 
collimated jet (Br1 and Bb1 CO knots from Fig.~\ref{fig:chanmapBeR}), which is not associated 
with the large-scale CO outflow direction and with the direction perpendicular to the A disk, but it is rather associated 
with the closest [O{\sc i}] 63~$\mu$m emission peaks.
               }
              \label{fig:COeOI}%
    \end{figure*}

In this research note, we present Atacama Large Millimeter/submillimeter Array (ALMA) Early Science Cycle 0 observations
of CO(2$-$1) towards VLA1623.
The goal is to study the physical properties of the jet emission in VLA1623 at high-angular resolution.


\section{Observations}
\label{sect:observations}

VLA1623 has been mapped in the CO(2$-$1) line with ALMA during the Early Science Cycle 0 
phase on April 8, 2012 (PI Murillo).
The details of the observations, calibration, and data reduction were presented in \citet{murillo2013b}.
Pointing coordinates were $\alpha_{\rm J2000}$=16$^{\rm h}$26$^{\rm m}$26$.\!\!^{s}$419 and 
$\delta_{\rm J2000}$=$-$24$^\circ$24$^{\prime}$29$.\!\!^{\prime\prime}$99. 
Observations were done in Band 6 (230 GHz) using the extended configuration, and they consisted 
of 16 antennae with a baseline range of $\sim$36$-$400~m for a total observing time of one hour.
The minimum baseline of 36~m yields a maximum recoverable scale of 4$^{\prime\prime}$
at the selected frequency, which is consistent with filtering most of the emission associated with the large-scale 
outflow (see Sect.~\ref{sect:discussion}).
The synthesised beam is 0$.\!\!^{\prime\prime}$74$\times$0$.\!\!^{\prime\prime}$58 (PA$=$93$^{\circ}$). 
The rms noise is 3$-$5~mJy~beam$^{-1}$ per 1~km~s$^{-1}$ channel.
Data was analysed using the GILDAS\footnote{\url{http://www.iram.fr/IRAMFR/GILDAS}} package. 

   \begin{figure*}
   \centering
   \includegraphics[width=0.98\textwidth]{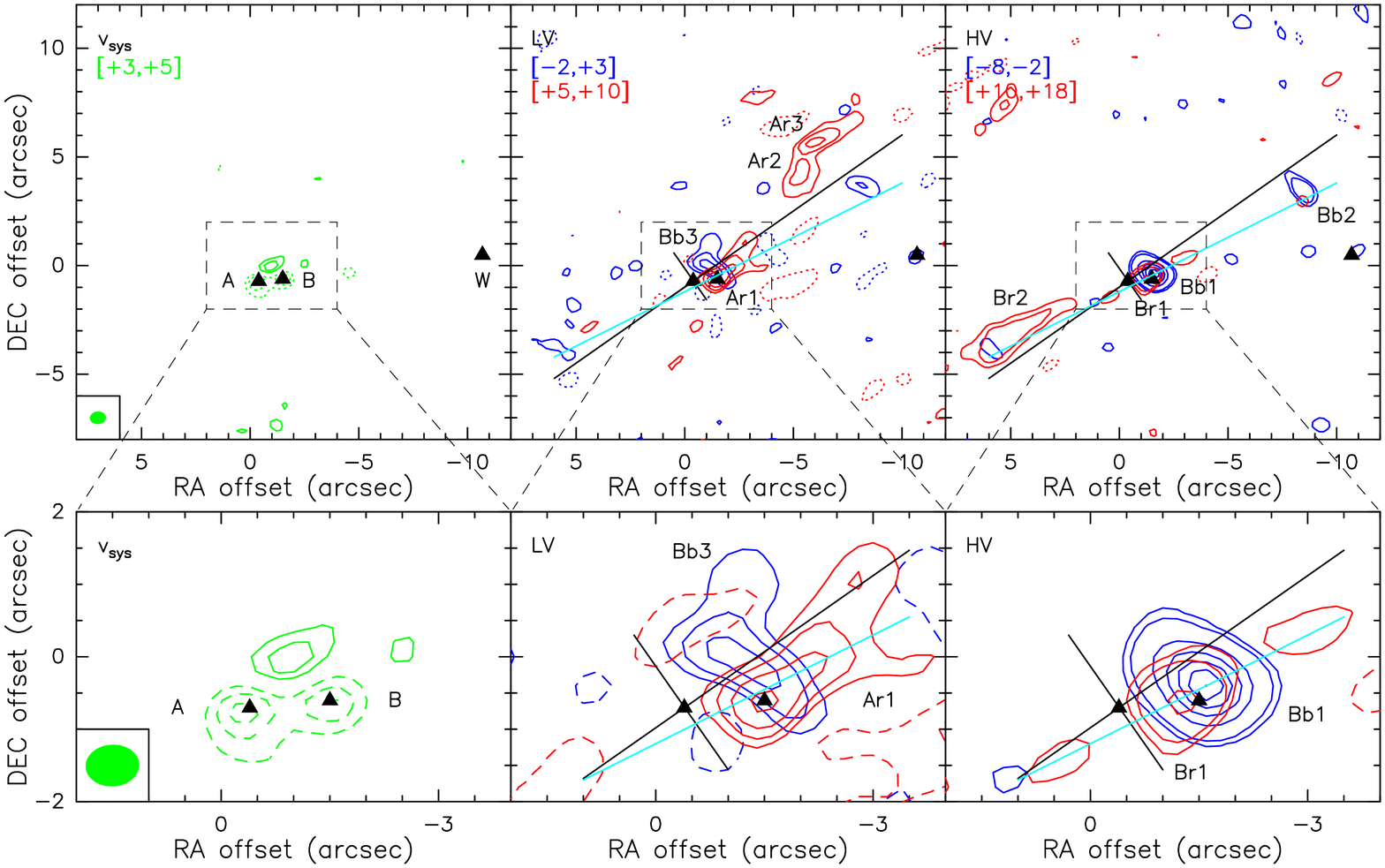}
   \caption{Channel maps of the ALMA CO(2$-$1) emission integrated over three velocity ranges: 
systemic velocity in the left panels ($v_{\rm sys}$$\pm$1~km~s$^{-1}$, where the systemic velocity is 3.6~km~s$^{-1}$, see
\citealt{yu1997,narayanan2006}), 
low-velocity in the middle panels (LV; $-$2, $+$3 km s$^{-1}$ and $+$5, $+$10 km s$^{-1}$), 
and high-velocity in the right panels (HV; $-$8, $-$2 km s$^{-1}$, and $+$10, $+$18 km s$^{-1}$). 
Offsets are given with respect to the pointing coordinates, i.e. $\alpha_{\rm J2000}$=16$^{\rm h}$26$^{\rm m}$26$.\!\!^{s}$419 and 
$\delta_{\rm J2000}$=$-$24$^\circ$24$^{\prime}$29$.\!\!^{\prime\prime}$99.
Contour levels of emission are traced at 3, 5, and 8~$\sigma$ levels and increase in steps of 5~$\sigma$, 
with the exception of the HV panels where they are shown at 3 and 5~$\sigma$ and increase in steps of 10~$\sigma$.
Negative contours are displayed with dotted (upper) and dashed (lower) lines 
starting at 3~$\sigma$ level and decreasing in steps of 3~$\sigma$.
The ALMA synthesised beam of the CO map is shown in the left panels.
Black triangles represent the three protostellar sources, which are labelled in the left panel. 
Other labels indicate the CO features discussed in the text.
Black and cyan solid lines are the same as in Fig.~\ref{fig:COeOI}.
               }
              \label{fig:chanmapBeR}%
    \end{figure*}


\section{Results}
\label{sect:results}

Figure~\ref{fig:COeOI} presents the comparison between the ALMA blue- and red-shifted CO(2$-$1) emission, 
the large-scale CO outflow emission, and the PACS [O{\sc i}] 63~$\mu$m line emission from \citet{nisini2015}.
The [O{\sc i}] 63~$\mu$m emission is extended along the outflow direction and shows several emission 
peaks corresponding to different shock episodes.
The direction perpendicular to the A disk (black solid line, PA=125$^{\circ}$) is consistent with the large-scale outflow direction 
(identified in Fig.~\ref{fig:COeOI} by the BLUE1/RED1 peaks) and with the two associated secondary [O{\sc i}] 63~$\mu$m peaks. 
Conversely, the closest [O{\sc i}] 63~$\mu$m peaks are not co-aligned with the A outflow direction, but they are instead 
aligned with source B (cyan line).

The ALMA data resolve the CO emission previously detected in CO(3$-$2) with SMA 
\citep[see][and Fig.~\ref{Afig:COchanmapSMA}, where the channel maps of the SMA observations are presented]{murillo2013a}; 
in particular, the southern (low-velocity) blue-shifted emission is almost completely filtered out by the ALMA observations. 
Moreover, no compact outflow emission seems to be associated with source W.  
Finally, the ALMA CO(2$-$1) emission is not associated with the large-scale outflow emission, 
and it possibly shows two outflow components: 
a collimated emission, identified by the northern blue- and southern red-shifted emissions along the cyan solid line, 
which is inversely oriented with respect to the large-scale outflow 
and associated with the closest [O{\sc i}] 63~$\mu$m emission peaks; and 
an arc-like feature of enhanced red-shifted CO intensity,
which surrounds the northern [O{\sc i}] 63~$\mu$m emission and may be associated with a cavity opened by the jet.

More clues about these components are provided by the channel maps of the red- and blue-shifted CO(2$-$1) emission 
presented in Figs.~\ref{fig:chanmapBeR} and \ref{Afig:COchanmap}.
The two components identified in the velocity-integrated emission (Fig.~\ref{fig:COeOI}) appear to be well 
separated in velocity, the red-shifted cavity-like structure dominating the emission at low projected velocities (Ar1, Ar2, and Ar3 knots), 
whereas the collimated jet associated with the closest [O{\sc i}] 63~$\mu$m peaks emitting in the higher-velocity channels 
(Br2, Br1, Bb1, and Bb2 knots).

Finally, compact blue- and red-shifted outflow emission at about 0$.\!\!^{\prime\prime}$3 from B are detected 
(Bb1 and Br1 knots).
We note that, assuming an inclination angle with respect to the line of sight of 75$^{\circ}$ as for the A outflow \citep{davis1999}, 
the maximum projected velocities of 14 km s$^{-1}$ 
\citep[with respect to the ambient systemic velocity of 3.6~km~s$^{-1}$,][]{yu1997,narayanan2006},
detected in association with the collimated jet component, 
correspond to de-projected velocities of 54 km s$^{-1}$. Even assuming an inclination angle of 45$^{\circ}$, 
de-projected velocities of about 20 km s$^{-1}$ are derived. Thus, we refer to the collimated jet component 
associated with the closest [O{\sc i}] 63~$\mu$m peaks and to the compact jet emission from B as fast jet components.


\section{Discussion}
\label{sect:discussion}

   \begin{figure*}
   \centering
   \includegraphics[width=0.7\textwidth]{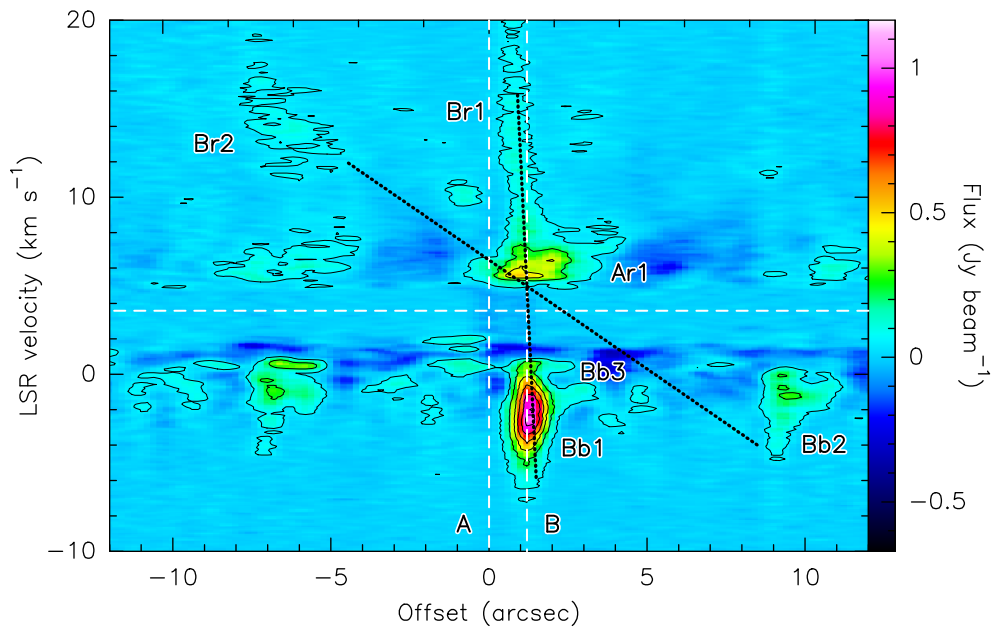}
   \caption{Position-velocity cut of the CO(2$-$1) emission in false colours and black contours
performed along the direction of the 
collimated jet identified by Br1 and Bb1 CO knots in Fig.~\ref{fig:chanmapBeR}
(along the cyan solid line in Figs.~\ref{fig:COeOI}$-$\ref{fig:chanmapBeR}, PA$=$116$^{\circ}$).
Labels are the same as in Fig.~\ref{fig:chanmapBeR}.
Dashed vertical white lines mark the positions of VLA1623A and VLA1623B, while the dashed horizontal white line 
indicates the ambient systemic velocity of 3.6 km s$^{-1}$, derived by \citet{narayanan2006} from 
N$_2$H$^+$ observations 
taken with the Five College Radio Astronomy Observatory (FCRAO) \citep[see also][]{yu1997}.
Dotted black lines connect the corresponding blue- and red-shifted knots associated with the 
possibly new fast and collimated jet emission detected with ALMA and labelled in Fig.~\ref{fig:chanmapBeR}. 
Both pairs of corresponding knots seem to be associated with B rather than with A.
               }
              \label{fig:PV}%
    \end{figure*}

We observed compact jet emission very close to B (Bb1 and Br1 knots).
In addition, farther from sources A and B we detect two outflow components: 
one faster red- and blue-shifted jet emission associated with the closest [O{\sc i}] 63~$\mu$m peaks and 
inversely oriented with respect to the large-scale outflow emission (Bb2 and Br2), 
and a slower northern red-shifted cavity-like structure surrounding the [O{\sc i}] 63~$\mu$m emission (Ar1, Ar2, and Ar3).
Two possible scenarios may explain the present observational results: 1) precession of the A jet, and 
2) the presence of a fast compact jet associated with source B.

In the first case, the misalignment of the closest [O{\sc i}] 63~$\mu$m emission peaks  with respect to the direction  
of the large-scale outflow 
may be explained with a scenario in which the A jet, with an inclination angle close to the plane of the sky, is precessing owing to the mutual interaction with the B companion. 
This causes a change in propagation direction of the outflow driven by A and the inversion of the blue- and red-shifted emissions.
We note that the presence of precessing jets in this region 
has already been suggested by \citet{caratti2006}, but on much larger scales than those probed by the present ALMA data.
In this view, the compact jet emission associated with the extremely young source B (Bb1 and Br1 knots) would represent a pole-on 
outflow, as already suggested by \citet{murillo2013a}.
However, the direction of the secondary, hence (in this view) of older [O{\sc i}] 63~$\mu$m peaks, 
is consistent with the direction perpendicular to the A disk (and with the large-scale outflow), 
whereas if the A jet was precessing
we would expect the direction of the closest [O{\sc i}] 63~$\mu$m peaks to be consistent with the direction of the A jet.

A second interpretation could be that the fast collimated jet emission associated with the 
closest [O{\sc i}] 63~$\mu$m peaks (Bb2 and Br2) and the high-velocity knots close to source B (Bb1 and Br1) 
represent a newly detected jet driven by B. 
This is supported by the position-velocity diagram along the direction of this fast collimated jet (identified by the Br1 and Bb1 knots), 
which is presented in Fig.~\ref{fig:PV}. It shows how the pairs of corresponding red- and blue-shifted knots 
lying along this possibly new jet seem rather to be associated with source B than with source A.
This would mean that the jet associated with B is not pole-on as previously argued by \citet{murillo2013a}, 
and it implies a very short dynamical timescale for the close compact emission knots (Br1 and Bb1).
In particular, assuming an inclination angle of 75$^{\circ}$ as for the A outflow \citep{davis1999}, 
the dynamical timescale of Br1 and Bb1, located at 
the projected distance from B of 0$.\!\!^{\prime\prime}$3 and moving at the radial velocity of $\sim$12 km s$^{-1}$
(Figs.~\ref{fig:chanmapBeR}$-$\ref{fig:PV}), is $\sim$4 yr.
Even assuming an inclination angle of 45$^{\circ}$ for the B jet, the dynamical timescale would be $\sim$14 yr.
%

On the other hand, the slower cavity-like structure would be associated with the large-scale outflow driven by A, which is 
possibly extended and poorly collimated and therefore almost completely filtered out by the ALMA observations.
We note that the lack of CO emission in our map near the systemic velocity is probably due to 
interferometric spatial filtering of extended emission from the envelope and the large-scale outflow, therefore our CO map can only trace emission from compact red- and blue-shifted material.
According to this scenario, source B is driving a much more compact and collimated outflow with respect to A.
In fact, if we assume for the B jet the same intrinsic length as observed for the A outflow, i.e. at least 60$-$100 arcsec,  
the projected length of the B jet of $\sim$8$^{\prime\prime}$ would imply an inclination for the B jet of $\sim$5$^{\circ}$$-$8$^{\circ}$, 
which can be excluded given the morphology of the jet emission and the well-separated jet lobes.
In conclusion, this interpretation is consistent with source B being younger than A, in a very early stage of evolution 
as already proposed by \citet{murillo2013a}, and driving a faster, more collimated, and more compact jet 
with respect to the large-scale slower outflow driven by A. 

Finally, the non-detection with ALMA of any outflow emission towards source W 
could confirm that source W is older than the other two sources, A and B \citep[][]{murillo2013a},
and thus has already swept out all surrounding material. 
Alternatively, the W outflow emission, if not compact, could be filtered out by the interferometer.

ALMA follow-up observations of optically thin (SO) and selective shock tracers (SiO) would be crucial for 
clarifying the proposed scenario.  
In particular, observations in both the extended and compact configurations are needed 
in order to avoid filtering of the large-scale emission and to eventually recover the emission from the large-scale outflow. 
This will allow the two possible jets to finally be disentangled and 
will put strong constraints on current theoretical models of multiple system formation.
Moreover, wide-band spectra of the A and B sources would be needed to 
understand their nature and chemical compositions (from the possible detection of complex organic molecule emission).


\section{Conclusions}
\label{sect:conclusions}

Compact jet emission is detected very close to source B.
In addition, two outflow components are detected further from sources A and B: 
one faster and more collimated jet emission associated with the closest [O{\sc i}] 63~$\mu$m peaks and 
inversely oriented with respect to the large-scale outflow emission; 
and a slower cavity-like structure surrounding the [O{\sc i}] 63~$\mu$m emission.
The observations seem to confirm 
that source B is younger than A, is in a very early stage of evolution, 
and is driving a faster, more collimated, and more compact jet, with respect to the large-scale slower outflow driven by A.
However, a different scenario of a precessing jet driven by A cannot be firmly excluded from the present observations.

\begin{acknowledgements}
This paper made use of the following ALMA data: ADS/JAO.ALMA 2011.0.00902.S and 2013.1.01004.S. ALMA is a partnership of ESO 
(representing its member states), NSF (USA), and NINS (Japan), together with NRC (Canada) and NSC and ASIAA (Taiwan), 
in cooperation with the Republic of Chile. The Joint ALMA Observatory is operated by ESO, AUI/NRAO, and NAOJ. 
The 2011.0.00902.S data were obtained by N.M.M. while she was a Masters student at National Tsing Hua University, Taiwan, 
under the supervision of S.P.L.
The work was partly supported by the ASI--INAF project 01/005/11/0 and 
the PRIN INAF 2012 -- Jets, disks, and the dawn of planets (JEDI). 
\end{acknowledgements}


\Online

\begin{appendix} 
\section{CO channel maps}
\begin{figure*}
\centering
\includegraphics[width=0.8\textwidth]{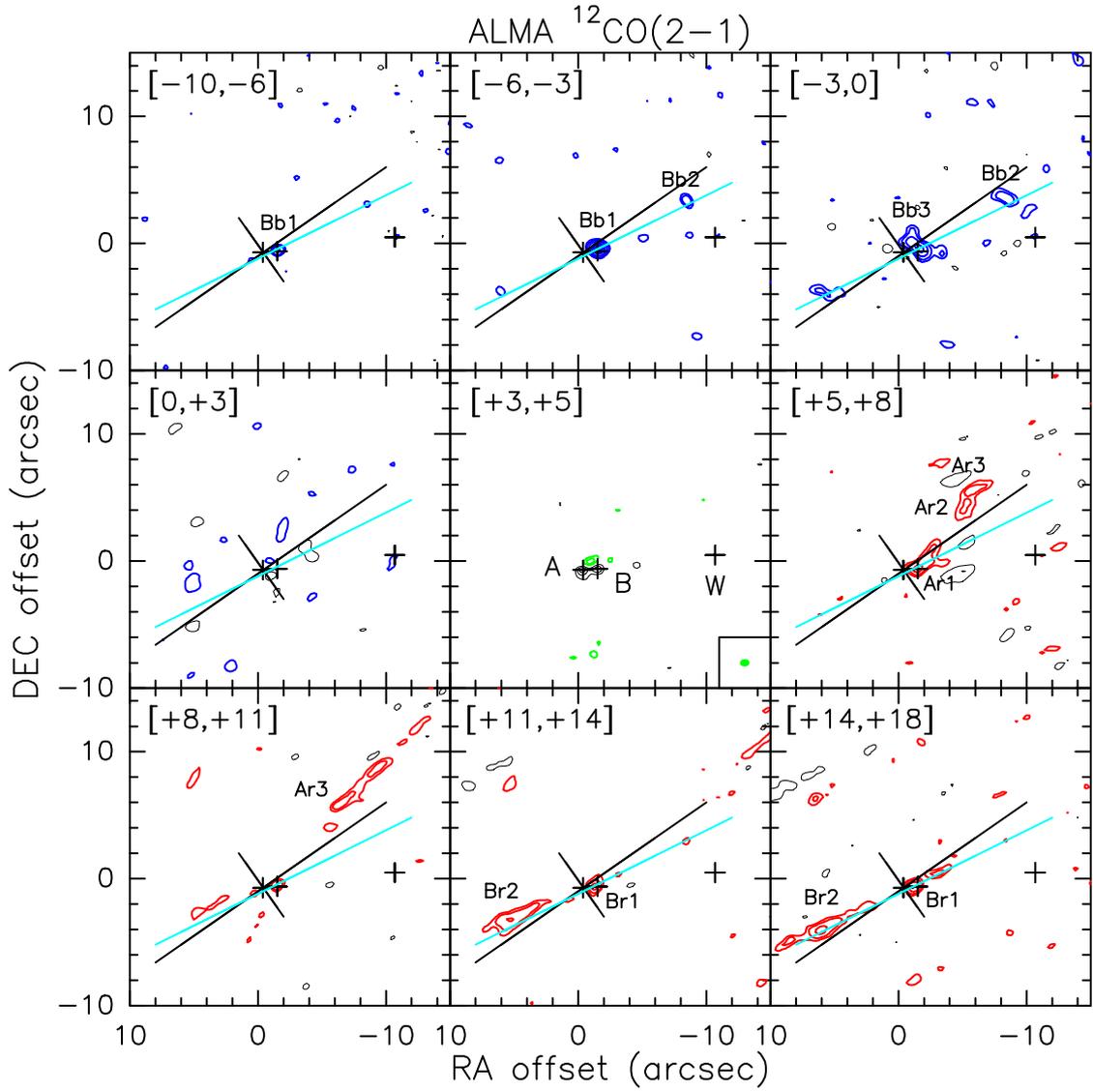}
\caption{Above: the channel maps of the ALMA CO(2$-$1) emission. 
The velocity range of each map is indicated in the top left corner.
The ALMA synthesised beam of the CO map is shown in the central panel.
Contour levels of emission are traced at 3, 5, and 10~$\sigma$ levels and increase in steps of 10~$\sigma$. 
Negative contours are displayed in black starting at 3~$\sigma$ level and decreasing in steps of 3~$\sigma$.
Black crosses represent the three protostars, which are labelled in the central panel. 
Other labels indicate the CO features discussed in the text (see also Fig.~\ref{fig:chanmapBeR}). 
Black and cyan solid lines are the same as in Fig.~\ref{fig:COeOI}.
The ALMA observations resolve the low-velocity blue-shifted emission (0, +3 km s$^{-1}$) detected with SMA
(see Fig.~\ref{Afig:COchanmapSMA}), 
which is possibly associated with extended emission and, therefore, filtered out by our high-resolution ALMA 
interferometric observations.
}
\label{Afig:COchanmap}
\end{figure*}

\begin{figure*}
\centering
\includegraphics[width=0.8\textwidth]{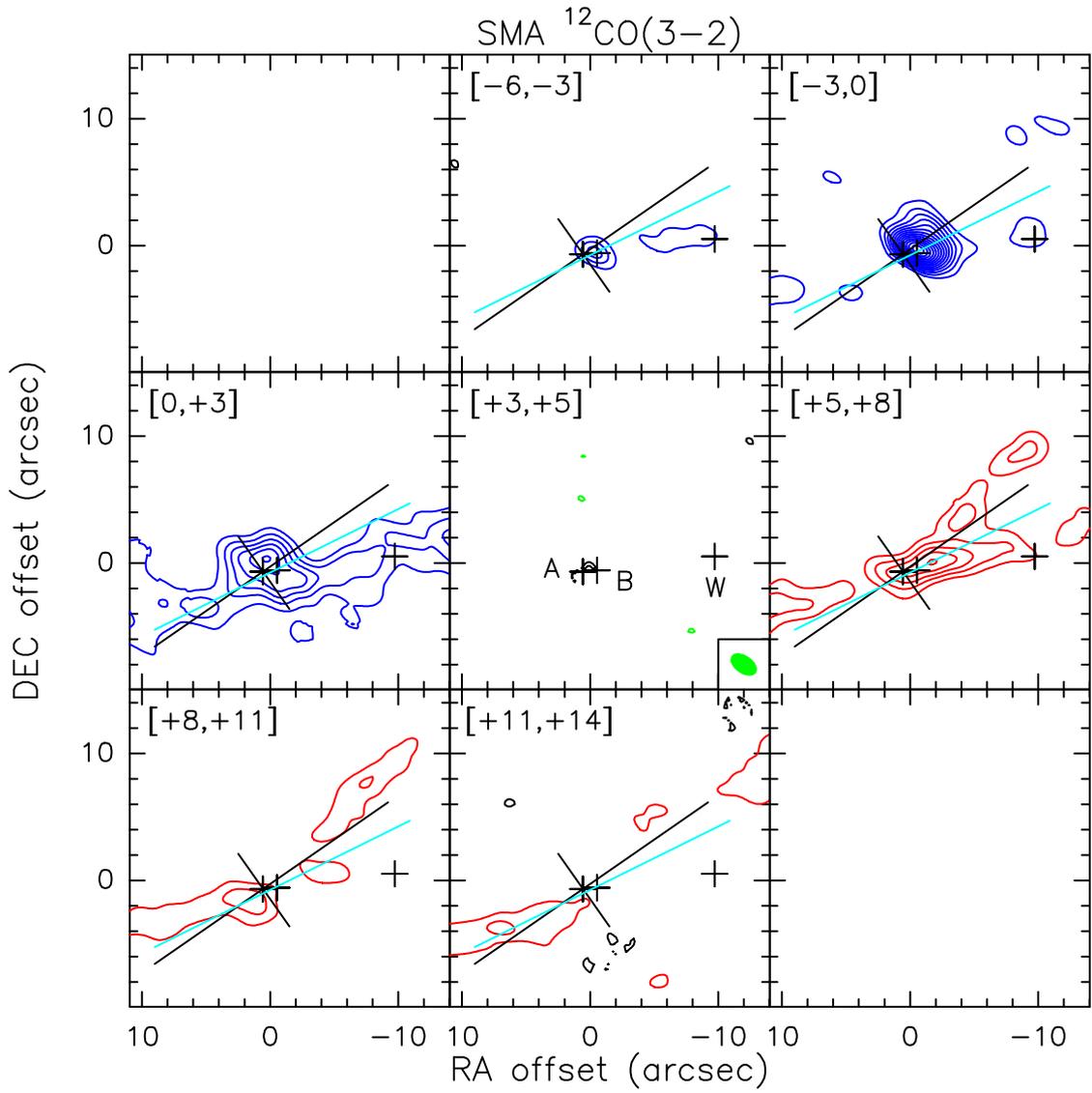}
\caption{Channel maps of the SMA CO(3$-$2) emission. 
The velocity range of each map is indicated in the top left corner.
The SMA synthesised beam of 2$.\!\!^{\prime\prime}$32$\times$1$.\!\!^{\prime\prime}$41 (PA$=$55$^{\circ}$) is shown in the central panel.
Contour levels of emission are traced starting at 3~$\sigma$ levels and increasing in steps of 3~$\sigma$. 
Negative contours are displayed in black starting at 3~$\sigma$ level and decreasing in steps of 3~$\sigma$.
Black crosses represent the three protostars, which are labelled in the central panel.
Black and cyan solid lines are the same as in Fig.~\ref{fig:COeOI}.
}
\label{Afig:COchanmapSMA}
\end{figure*}

We show in Fig.~\ref{Afig:COchanmap} the channel maps of the blue- and red-shifted ALMA CO(2$-$1) (continuum subtracted) 
emission towards VLA1623, which are analysed and discussed in the main text (see Sect.~\ref{sect:results}).
The images clearly show two gas components that are well separated in velocity: a red-shifted cavity-like structure dominating the emission 
at low projected velocities (Ar1, Ar2, and Ar3 knots); and 
a collimated emission in the higher-velocity channels (Br2, Br1, Bb1, and Bb2 knots along the cyan line) and 
associated with the closest [O{\sc i}] 63~$\mu$m peaks (see Fig.~\ref{fig:COeOI}).
In particular, compact high-velocity outflow emission is detected close to source B (Bb1 and Br1 knots).
The channel maps of the blue- and red-shifted SMA CO(3$-$2) emission are also presented in Fig.~\ref{Afig:COchanmapSMA} 
for comparison.
We note that the low-velocity blue-shifted emission between 0 and $+$3~km~s$^{-1}$ is resolved by the ALMA observations.

\end{appendix}

\end{document}